# Single-layer magnet phase in intrinsic magnetic topological insulators, [MnTe][Bi$_2$Te$_3$]$_n$, far beyond the thermodynamic limit

*Deepti Jain[1]\*, Hee Taek Yi[1], Xiong Yao[1,6,a], Alessandro R. Mazza[2,3], An-Hsi Chen[2], Kim Kisslinger[4], Myung-Geun Han[5], Matthew Brahlek[2] and Seongshik Oh[1,6]\**

[1] Department of Physics and Astronomy, Rutgers, The State University of New Jersey, Piscataway, NJ 08854, USA

[2] Materials Science and Technology Division, Oak Ridge National Laboratory, Oak Ridge, TN 37831, USA

[3] Present address: Materials Science and Technology Division, Los Alamos National Laboratory, Los Alamos, New Mexico 87545, USA

[4] Center for Functional Nanomaterials, Brookhaven National Laboratory, Upton, NY 11973, USA

[5] Condensed Matter Physics and Materials Science, Brookhaven National Laboratory, Upton, NY 11973, USA

[6] Center for Quantum Materials Synthesis, Rutgers, The State University of New Jersey, Piscataway, NJ 08854, USA

[a] Present address: Ningbo Institute of Materials Technology and Engineering, Chinese Academy of Sciences, Ningbo 315201, China

ABSTRACT

The intrinsic magnetic topological insulator (IMTI) family $[MnTe][Bi_2Te_3]_n$ has demonstrated magneto-topological properties dependent on $n$, making it a promising platform for advanced electronics and spintronics. However, due to technical barriers in sample synthesis, their properties in the large $n$ limit remain unknown. To overcome this, we utilized the atomic layer-by-layer molecular beam epitaxy (ALL-MBE) technique and achieved IMTIs with $n$ as large as 15, far beyond the previously reported in bulk crystals or thin films. Then, we discover that the "single-layer magnet (SLM)" phase, primarily determined by intralayer ferromagnetic coupling, emerges for $n >$ ~4 and remains little affected up to $n = 15$. Nonetheless, still, non-zero, interlayer ferromagnetic coupling is necessary to stabilize the SLM phase, suggesting that the SLM phase eventually disappears in the $n \to \infty$ limit. This study uncovers the secrets of IMTIs beyond the thermodynamic limit and opens a door to diverse magneto-topological applications.

[MnTe][Bi$_2$Te$_3$]$_n$, the only intrinsic magnetic topological insulators (IMTIs) known to exist so far, is a stoichiometric material system that is topologically non-trivial with an inherent magnetic order. These unique attributes make it an ideal platform to host a myriad of topological states such as axion insulators, magnetic Weyl semimetals, Chern insulators, topological magnetoelectric effect, and high temperature quantum anomalous Hall effects.[1-9] Moreover, their built-in magnetism and naturally-ordered structure ensures more homogenous properties as compared to magnetically doped topological insulators, which is beneficial for reproducibility and scalability in practical applications.

Depending on *n,* i.e. the number of Bi$_2$Te$_3$ layers, the magnetic behavior of [MnTe][Bi$_2$Te$_3$]$_n$ varies. In the first member of this family of materials, MnBi$_2$Te$_4$ ($n = 1$), each unit cell can be viewed as a layer of MnTe, an antiferromagnet, inserted within topological insulator Bi$_2$Te$_3$. Together they stack as Te-Bi-Te-Mn-Te-Bi-Te, forming self-organized septuple layers (SLs) with each SL bonded to the next via weak van der Waals forces. This results in a magnetic structure akin to an A-type antiferromagnet; ferromagnetic (FM) order within each SL and antiferromagnetic (AFM) coupling between adjacent SLs.[6, 10, 11] By inserting ($n - 1$) Bi$_2$Te$_3$ units between each SL, the other members of the [MnTe][Bi$_2$Te$_3$]$_n$ family can be realized, as seen in the schematic of Figure 1a. As *n* increases, the AFM interlayer exchange coupling (IEC) between SLs is weakened, and the magnetic properties of [MnTe][Bi$_2$Te$_3$]$_n$ can be tuned. Accordingly, the IEC progressively decreases in MnBi$_4$Te$_7$ ($n = 2$) and MnBi$_6$Te$_{10}$ ($n = 3$). Some studies report that $n = 2$ and $n = 3$ have a weakened AFM order (as compared to $n = 1$),[12-15] while others note competing AFM and FM orders.[16, 17] Ultimately, the IEC completely vanishes for MnBi$_8$Te$_{13}$ ($n = 4$) resulting in a FM state.[12, 18] For $n \geq 4$, once the individual FM SLs are decoupled, a "single-layer magnet (SLM)" phase has been predicted in which the SLs behave like independent 2D ferromagnets.[12, 19] In other words, it is expected that in the $n \to \infty$ limit, the FM ordering will persist. Unfortunately, this prediction of SLM phase has not been verified beyond $n = 5$ due to challenges in synthesizing high order $n$ members.

To date, the most extensively studied members of the $[MnTe][Bi_2Te_3]_n$ material system are bulk crystals and exfoliated flakes of $n$ = 1, 2, 3 and 4. This is because the temperature range in which $[MnTe][Bi_2Te_3]_n$ compounds crystallize gets narrower as $n$ increases. Hence, it becomes progressively more difficult to obtain high purity crystals for higher order $n$ that are of adequate size.[20, 21] As a result, growth of only $n \leq 7$ phases has been reported. Here, however, we demonstrate that these thermodynamic limitations can be bypassed with the atomic-layer-by-layer molecular beam epitaxy (ALL-MBE) technique, and present previously-unknown electronic and magnetic properties of these materials well beyond the bulk limit.

Figure 1b shows a schematic for the growth structure on $Al_2O_3(0001)$ substrates. To improve sticking to the substrate, a buffer layer of ~1 nm-thick $Cr_2O_3$ is introduced between the substrate and $[MnTe][Bi_2Te_3]_n$ films, similar to our previous works (see complete details in the Methods section).[22-25] Next, depending on the choice of $n$, each individual layer of MnTe and $Bi_2Te_3$ is deposited sequentially at 260-300 °C. This is achieved by opening and closing the elemental Mn and Bi shutters, while the Te shutter is kept open, with ~6-7 times higher Te flux than Bi and Mn. This growth approach is based on the fact that the septuple layer, $MnBi_2Te_4$, is thermodynamically more stable than having a heterostructure of two separate layers of MnTe and $Bi_2Te_3$ on top of each other. Accordingly, when a MnTe layer is deposited between $Bi_2Te_3$ layers, it naturally forms a septuple layer of $MnBi_2Te_4$, while additional $Bi_2Te_3$ layers form the quintuple layers below and above the $MnBi_2Te_4$ layer. This way, we are able to "pin" the Mn atoms and confine the SLs in the desired positions: this mode of growth is critical to achieving the high order

$[MnTe][Bi_2Te_3]_n$ structures reported here. Another significant advantage of such a layer-by-layer approach compared to the more common co-deposition method is that it is not critical to control the Mn:Bi ratio. As shown in the schematic of Figure 1b, the film is grown in "blocks", each comprising of a MnTe layer and $n$ QLs of $Bi_2Te_3$. Each block is repeated $m$ times to obtain the desired thickness. In the samples discussed below, most samples are uncapped while some are capped with Te, but their properties are not significantly affected due to this discrepancy, especially the ones with high order $n$. All the films discussed in Figures 3 and 4 have $m = 10$, except for the $n = 1$ sample, which has $m = 32$.

The crystallinity and 2D surfaces of our films are corroborated by reflection high energy electron diffraction (RHEED) images taken *in situ* after growth (Figure S3). Figure 2a shows the x-ray $2\theta$ scans for $n = 1, 3, 5, 7$ and $9$ $[MnTe][Bi_2Te_3]_n$ films, including peaks belonging to the $Al_2O_3$ substrate and additional peaks due to Te capping in some films. For $n = 15$, i.e. the sample with the largest $n$, due to the extremely long periodicity ($> 30$ nm), the peaks in the XRD pattern are so close to each other that they are included in the Supporting information, to distinguish the peaks better (Figure S1). Below the XRD scan for each film, the expected values of the (00$l$) reflections based on the $c$-axis lattice constant of each $[MnTe][Bi_2Te_3]_n$ have been marked as open triangles. The patterns for $MnBi_2Te_4$, and $MnBi_6Te_{10}$ agree with their counterparts in multiple references.[14, 26] Due to lack of existence of concrete XRD data for higher order films beyond $n = 4$, their peaks could not be compared with any references, but they match well with their expected values, indicating the formation of ordered structures of $[MnTe][Bi_2Te_3]_n$ along the (0001) direction. High-angle annular dark-field scanning transmission electron microscopy (HAADF-STEM) measurements were carried out on one of the samples: $MnBi_{14}Te_{22}$. Structurally, $MnBi_{14}Te_{22}$ should comprise 6 QLs of $Bi_2Te_3$ inserted between each $MnBi_2Te_4$ SL. These bands of SLs and QLs can be seen in the STEM image in Figure 2b, along with a sharp interface between the film and Te capping, implying a highly ordered growth. The complete STEM image can also be seen in Figure S2. It should be noted here that even though we expect 3 QL of $Bi_2Te_3$ between the Te capping layer and the topmost $MnBi_2Te_4$ septuple layer for the $[MnTe][Bi_2Te_3]_7$ structure based on our growth sequence, the STEM image shows only 2 QL of $Bi_2Te_3$. This implies that Mn/Bi atoms can diffuse vertically over 1~2 nanometers during the growth. Nonetheless, the overall superlattice periods both in XRD and in STEM are consistent with the designed superlattice order $n$.

The magnetic transition temperatures for $n = 1, 2, 3$ and 5 as observed from their temperature-dependent sheet resistance have been highlighted in Figure 3a. All samples exhibit metallic behavior. The metallicity also increases with the order of $n$, which is not surprising due to the addition of $Bi_2Te_3$ QLs in each subsequent order. Each of the kinks marked corresponds to the respective magnetic transition temperature. These temperatures are approximately 24 K, 14 K, 11 K and 9 K for $n = 1, 2, 3$ and 5 respectively, which are in good agreement with their bulk and MBE-grown counterparts.[11-15, 27, 28] The type of magnetic transition is better understood from the magneto-transport measurements ($B//c$, 2 K) of these samples. Figures 3b,d,f,h show the Hall resistance, with the linear part (for $B > 5$ T) subtracted in order to better visualize their magnetic properties. Meanwhile, Figures 3c,e,g,i illustrate the magnetic field dependence of the normalized sheet resistance $R_N = (R-R_{min})/(R_{max}-R_{min})$ with $R_{max}$ and $R_{min}$ being the maximum and minimum resistance respectively, allowing us to compare all the samples on the same scale. For $n = 1$ (Figure 3c), an "M shaped" curve is observed with peaks at $\pm 3.5$ T, consistent with a spin-flop transition seen in multiple reports.[9, 29, 30] The shape of the curve is slightly different from those of bulk crystals, in which there is a plateau instead of a "U shape" in between the two spin-flop peaks.

However, it is similar to the form observed in other MBE-grown films and gate-controlled exfoliated flakes.[9, 30-33] In Figure 3b, the prominent spike-like features in the Hall resistance coincide with the spin-flop transitions in Figure 3c. We also see a weak hysteretic (FM-like) response in some $n = 1$ films, possibly arising from Mn doping of the $Bi_2Te_3$ QLs.[28, 30, 34, 35] For $n = 2$, in Figure 3d,e, drastically different properties emerge, as the IEC decreases. Instead of a spin-flop, the magnetoresistance curve resembles the signature butterfly shape corresponding to FM order (Figure 3e). We also see two small peaks at ±2.5 T which are likely due to spin-flop of isolated $n = 1$ SLs in the sample.[28, 36] The Hall resistance of $n = 2$ (Figure 3d), yields a hysteresis which looks quite different from that of a typical ferromagnet. This is the characteristic of a spin-flip transition when the interlayer AFM coupling becomes weaker and smaller than the magnetic anisotropy, resulting in a FM-like hysteresis at low temperatures.[16, 37] The coercive field of this hysteresis, $B_c = 0.13$ T is consistent with the spin-flip field of bulk $MnBi_4Te_7$.[14-16, 27] Similarly, $B_c = 0.08$ T for $n = 3$ (Figure 3f), matches well with the spin-flip field of $MnBi_6Te_{10}$.[12, 16, 27] One can see a gradual change of the Hall resistance in the insets of Figures 3d,f,h as $n$ increases, with $n = 5$ looking the most like a hysteresis loop typical for FM materials, having $B_c = 0.05$ T. This is again comparable to previously studied $MnBi_{10}Te_{16}$ and $MnBi_8Te_{13}$,[12, 38] the compounds in which the interlayer AFM coupling ceases to exist.

      Figure 4 presents the temperature and magnetic field dependence of the Hall resistance of the remaining samples, $n = 7, 9, 11$ and $15$ that are predicted to be FM: it is notable that this series ($n \geq 7$) has never been demonstrated before, neither in bulk nor in thin films. The corresponding field dependence of sheet resistance can be seen in Figure S4. When non-FM materials are cooled under zero magnetic field, the Hall resistance should ideally be zero for all temperatures. But in the case of FM materials, an anomalous Hall signal starts to develop below $T_c$, i.e. when

spontaneous magnetization overcomes the thermal fluctuations. Such a signal is observed in the temperature-dependent Hall resistance curves for zero-field-cooled $n$ = 7, 9, 11 and 15 samples, which are offset in Figure 4a to easily compare the transition temperatures. Remarkably, clear transitions are seen around the same temperature, ~10 K, for each $R_{xy}$(T) curve. Magnetic field sweeps of the Hall resistance of these samples at 2 K yield hysteresis loops confirming the FM order (Figure 4b). Another common feature among all the samples is that that their coercive fields, $B_c$, are all very similar to ~0.05 T. Both $T_c$ and $B_c$ of $n$ = 7, 9, 11 and 15 are similar to those observed for $n$ = 5, indicating that all phases for $n \geq 5$ have the same FM properties. This is possible only if these properties arise solely from the intralayer FM interactions within each SL, and any interlayer coupling is absent (or negligible). On comparing our results with existing literature and theoretical predictions, we conclude that $n \geq 4$ indeed forms the SLM phase. Such a phase has robust $T_c \approx 10$ K and $B_c \approx 0.05$ T, completely independent of the number of $Bi_2Te_3$ layers between each magnetic MnTe layer and is here experimentally proven to survive at least up to $n$ = 15.

As mentioned before, all the samples in Figure 4 possess the structure shown in Figure 1b, with each "block" repeated 10 times. Our ability to engineer these structures layer-by-layer allows us to grow them with desired number of block repetitions. In the following section we present a systematic study of Hall resistance of $n$ = 5 and $n$ = 15 samples, the two extremes of the SLM phase in our study, for different number of block repetitions, $m$. In $R_{xy}$(T) of Figure 5c, $T_c$ of ~10 K is observed for all $m$ in $n$ = 5, but the corresponding $B_c \approx 0.05$ T in Figure 5d, is seen only for $m$ = 2 and $m$ = 3, and $B_c$ becomes almost negligible for $m$ = 1. On the other hand, in $n$ = 15, $m$ = 1 and 2 show no signs of ferromagnetism. Nonetheless, a transition temperature ($T_c \approx 10$ K) and hysteresis ($B_c \approx 0.05$ T) appear quite abruptly for $m$ = 3 and remain almost unchanged until $m$ = 10 (Figures 5a,b). For $m$ = 1, SLM phase signatures are seen in neither $n$ = 5 nor $n$ = 15. The FM hysteresis for $m = 1$ in $n = 5$ is very small and could be due to Mn doping of the $Bi_2Te_3$ layers, unrelated to 1 SL ferromagnetism. Similar absence of FM signal has been observed in 1 SL of a $MnBi_4Te_7$ thin flake sample,[39] and has been ascribed to the infeasibility of long-range order in 2D layers as described by the Mermin-Wagner theorem.[40] The values of $T_c$ and $B_c$ for $m \geq 2$ in $n = 5$ and $m \geq 3$ in $n = 15$, are similar to those observed in the samples in Figure 4 and can be attributed to the SLM phase. Interestingly, for $m = 2$, while FM features appear in $n = 5$, there are none in $n = 15$. Ideally, the SLM phase should show up in $m = 2$ for both $n = 5$ and $n = 15$, but its signatures appear at a higher $m$ value for $n = 15$ as compared to $n = 5$. This suggests that there is a weak, yet non-zero, long-range interaction between each SL that evidently gets weaker as more $Bi_2Te_3$ QLs are inserted in between. Since the FM features appear after a minimum number of $m$ for $n = 15$, we can also infer that this weak interaction is necessary to stabilize the overall FM order. Furthermore, the almost constant values of $T_c$ and $B_c$ imply that this interlayer FM interaction is much weaker than the intralayer super-exchange FM coupling. We speculate that this weak interlayer FM coupling could be due to either a carrier-mediated RKKY interaction or a dipole interaction between the FM SLs or both. Multiple studies indicate that defects such as antisite defects and Mn vacancies enhance interlayer FM coupling.[41-45] Hence, based on our results and the role of such defects in $[MnTe][Bi_2Te_3]_n$, artificially introducing small amount of defects could also help stabilize FM ordering, particularly in high order $n$.

In summary, by utilizing the ALL-MBE technique, we have synthesized the $[MnTe][Bi_2Te_3]_n$ series far beyond the previously reported, and revealed their hidden properties in the large $n$ limit. The FM signatures for the SLM phase ($n \geq 5$) are found to be independent of the Mn-Mn distance with $T_c$ and $B_c$ values of ~10 K and ~0.05 T, respectively, and survive up to the highest order that we grew, $n = 15$. This observation implies that the FM order is almost entirely determined by the strong intralayer FM interactions within each SL. Nonetheless, the thickness-dependent studies suggest that there should still exist a non-zero interlayer FM coupling between SLs, to stabilize the FM order. Hence, we can conclude that as $n \to \infty$, the [MnTe][Bi$_2$Te$_3$]$_n$ system does not exhibit ferromagnetism. Although the exact origin of this interlayer FM interaction is debatable and requires further in-depth studies, our work demonstrates that there are many untapped natures in the [MnTe][Bi$_2$Te$_3$]$_n$ family of materials that are hidden beyond the thermodynamic limit.

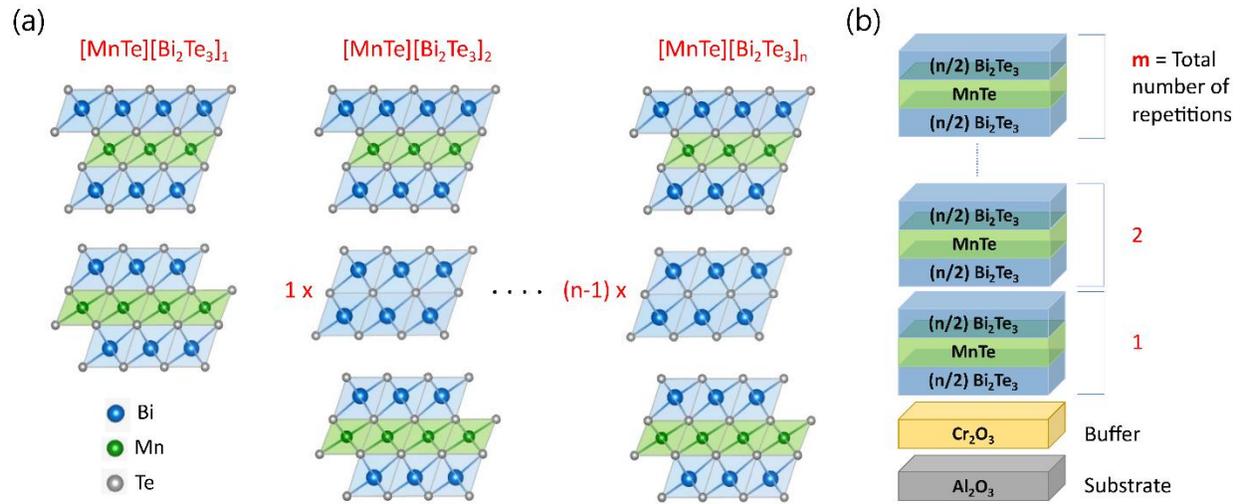

Figure 1. Crystal structure and growth schematic of [MnTe][Bi$_2$Te$_3$]$_n$. (a) Illustration of the generalized structure for all [MnTe][Bi$_2$Te$_3$]$_n$, with the number of Bi$_2$Te$_3$ QLs in between two SLs increasing with each order. (b) A schematic depicting the layer-by-layer growth of [MnTe][Bi$_2$Te$_3$]$_n$ on Al$_2$O$_3$(0001) substrates. Each "block" contains one MnTe layer inserted between $n$ Bi$_2$Te$_3$ QLs, which is then repeated $m$ times depending on the required thickness or total number of SLs.

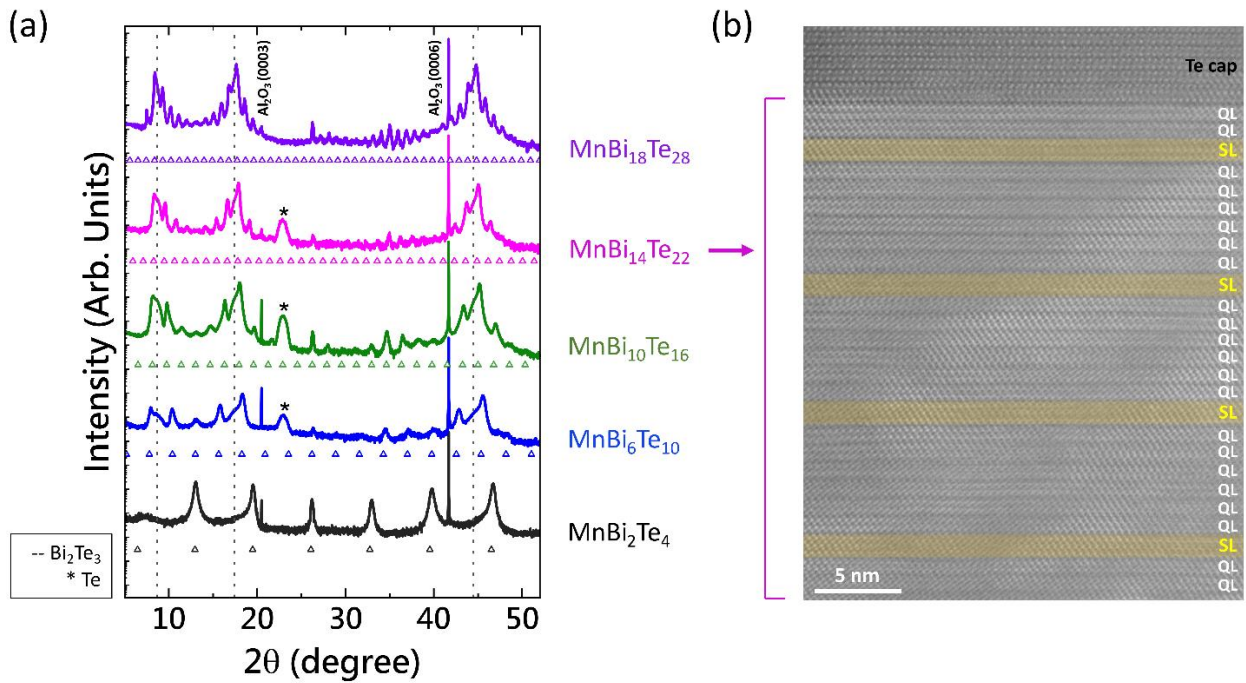

Figure 2. X-Ray Diffraction patterns and STEM. (a) XRD patterns of [MnTe][Bi$_2$Te$_3$]$_n$ for $n$ = 1, 3, 5, 7 and 9 with their expected peak values represented by open triangles below each pattern. Additionally, Al$_2$O$_3$ (substrate), Te (capping) and Bi$_2$Te$_3$ peaks have also been labelled and marked. (b) Cross-sectional HAADF-STEM image of MnBi$_{14}$Te$_{22}$ ($n$ = 7) illustrating the MnBi$_2$Te$_4$ SLs (marked as yellow bands to guide the eye) separated by 6 QLs of Bi$_2$Te$_3$.

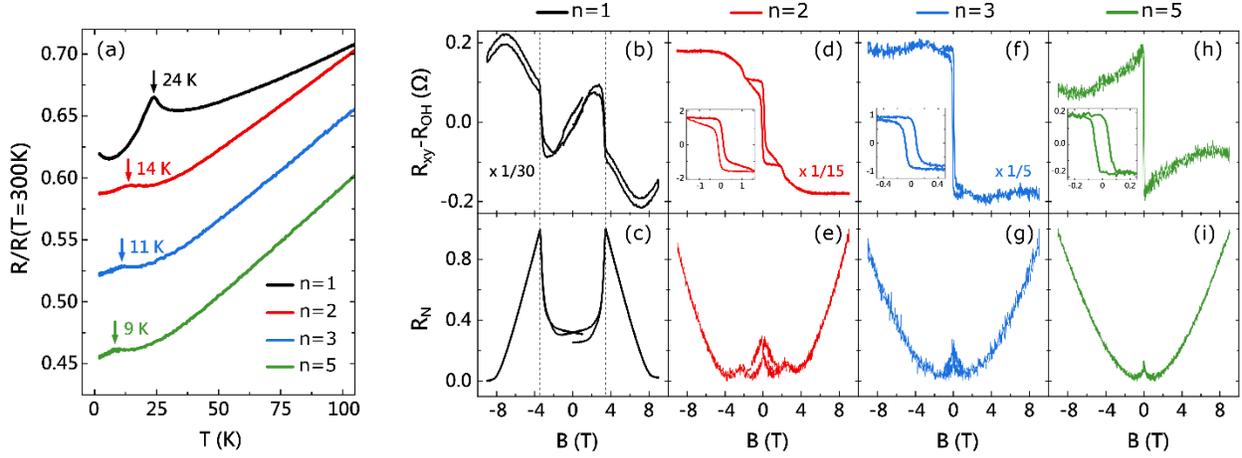

Figure 3. Transport properties of $n$ = 1, 2, 3 and 5. (a) Normalized sheet resistance vs temperature plot for $n$ = 1, 2, 3 and 5. The magnetic transition temperature for each curve is highlighted. (b,d,f,h) Field dependence of Hall resistance after subtracting ordinary Hall effect in the region $B > 5$ T, for $n$ = 1, 2, 3 and 5 respectively. (c, e, g, i) Field dependence of normalized sheet resistance $R_N = (R-R_{min})/(R_{max}-R_{min})$ for $n$ = 1, 2, 3 and 5 respectively. All measurements were taken at 2 K with $B//c$.

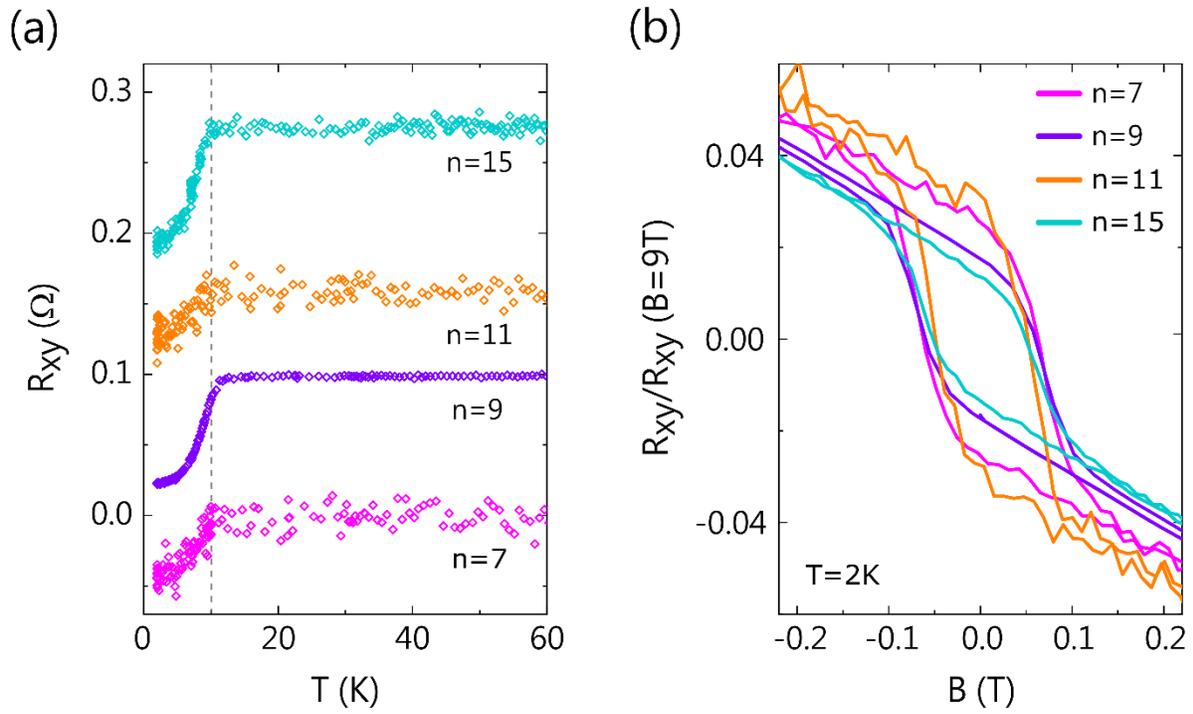

Figure 4. Hall resistance for *n* = 7, 9, 11 and 15. (a) Zero-field-cooled temperature dependence of Hall resistance for *n* = 7, 9, 11 and 15. The curves are offset with respect to each other. (b) Field dependence of normalized Hall resistance for *n* = 7, 9, 11 and 15 at 2 K with *B*//*c*.

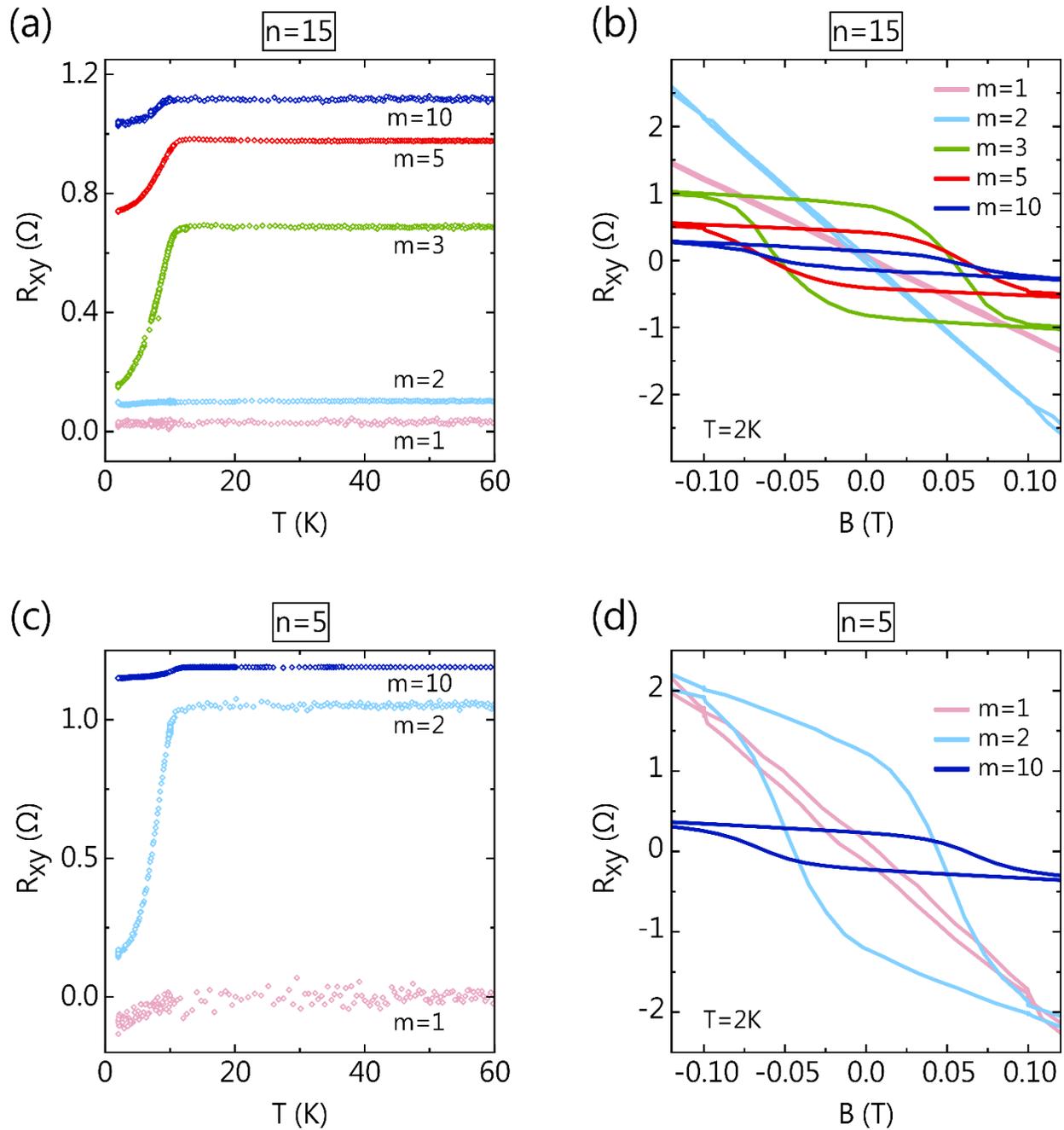

Figure 5. Thickness dependence of the Hall resistance for the SLM phase. Temperature and field dependence at 2 K of (a,b) $n = 15$ and (c,d) $n = 5$ for different values of $m$, where $m$ is the number of "block" repetitions in Figure 2.

ACKNOWLEDGEMENTS

This work is supported by Army Research Office's W911NF2010108 and MURI W911NF2020166. The work at Oak Ridge National Laboratory is supported by the U. S. Department of Energy (DOE), Office of Science, Basic Energy Sciences (BES), Materials Sciences and Engineering Division. The scanning transmission electron microscopy work performed at Brookhaven National Laboratory is sponsored by the US Department of Energy, Basic Energy Sciences, Materials Sciences and Engineering Division, under contract no. DE-SC0012704. This research used the Electron Microscopy resources (the Helios G5 FIB) of the Center for Functional Nanomaterials (CFN), which is a U.S. Department of Energy Office of Science User Facility, at Brookhaven National Laboratory under Contract No. DE-SC0012704. We acknowledge Hussein Hijazi for RBS measurements.

Supporting Information

# Single-layer magnet phase in intrinsic magnetic topological insulators, [MnTe][Bi$_2$Te$_3$]$_n$, far beyond the thermodynamic limit

*Deepti Jain[1]\*, Hee Taek Yi[1], Xiong Yao[1,6,a], Alessandro R. Mazza[2,3], An-Hsi Chen[2], Kim Kisslinger[4], Myung-Geun Han[5], Matthew Brahlek[2] and Seongshik Oh[1,6]\**

[1] Department of Physics and Astronomy, Rutgers, The State University of New Jersey, Piscataway, NJ 08854, USA

[2] Materials Science and Technology Division, Oak Ridge National Laboratory, Oak Ridge, TN 37831, USA

[3] Present address: Materials Science and Technology Division, Los Alamos National Laboratory, Los Alamos, New Mexico 87545, USA

[4] Center for Functional Nanomaterials, Brookhaven National Laboratory, Upton, NY 11973, USA

[5] Condensed Matter Physics and Materials Science, Brookhaven National Laboratory, Upton, NY 11973, USA

[6] Center for Quantum Materials Synthesis, Rutgers, The State University of New Jersey, Piscataway, NJ 08854, USA

[a] Present address: Ningbo Institute of Materials Technology and Engineering, Chinese Academy of Sciences, Ningbo 315201, China

\*Corresponding authors' email: jain@physics.rutgers.edu, ohsean@physics.rutgers.edu

**Experimental Methods**

**Thin film growth**: All films were grown on $10 \times 10 \times 0.5$ mm$^3$ Al$_2$O$_3$ (0001) substrates using a custom-built MBE system (SVTA) with base pressure of ~$10^{-10}$ Torr. The substrates were cleaned *ex situ* with UV generated ozone followed by *in situ* heating up to 750 °C under oxygen pressure of $1 \times 10^{-6}$ Torr. The Cr$_2$O$_3$ buffer layer was deposited at 700 °C under oxygen pressure of $1 \times 10^{-6}$ Torr, after which the substrate was cooled down to the growth temperature of [MnTe][Bi$_2$Te$_3$]$_n$. The sources used were high purity (99.999%) elemental Bi, Mn, Te, and Cr which were thermally evaporated using standard effusion cells. Source fluxes were calibrated *in situ* with a quartz crystal micro-balance and *ex situ* with Rutherford backscattering spectroscopy.

**Transport measurements**: The samples were prepared for transport measurement by using manually pressed indium wires in van der Pauw geometry. Magnetoresistance and Hall resistance measurements were carried out in a Physical Property Measurement System (PPMS, Quantum Design inc.) down to 2 K. Keithley 2400 source-measure unit and 7001 switch matrix system, controlled by a LabView program were used to gather data.

**XRD and STEM**: XRD was carried out, using a Panalytical X'Pert Pro and a monochromated Cu K$_{\alpha1}$ source. The STEM sample was prepared using a FEI Helios G5 UX focused ion beam system with final Ga+ milling performed at 2 keV. Then, the HAADF-STEM was performed with a JEOL ARM 200CF equipped with a cold field emission gun and spherical aberration correctors, which was operated at 200 kV. The detection angles for HAADF imaging were ranging from 68 to 280 mrad.

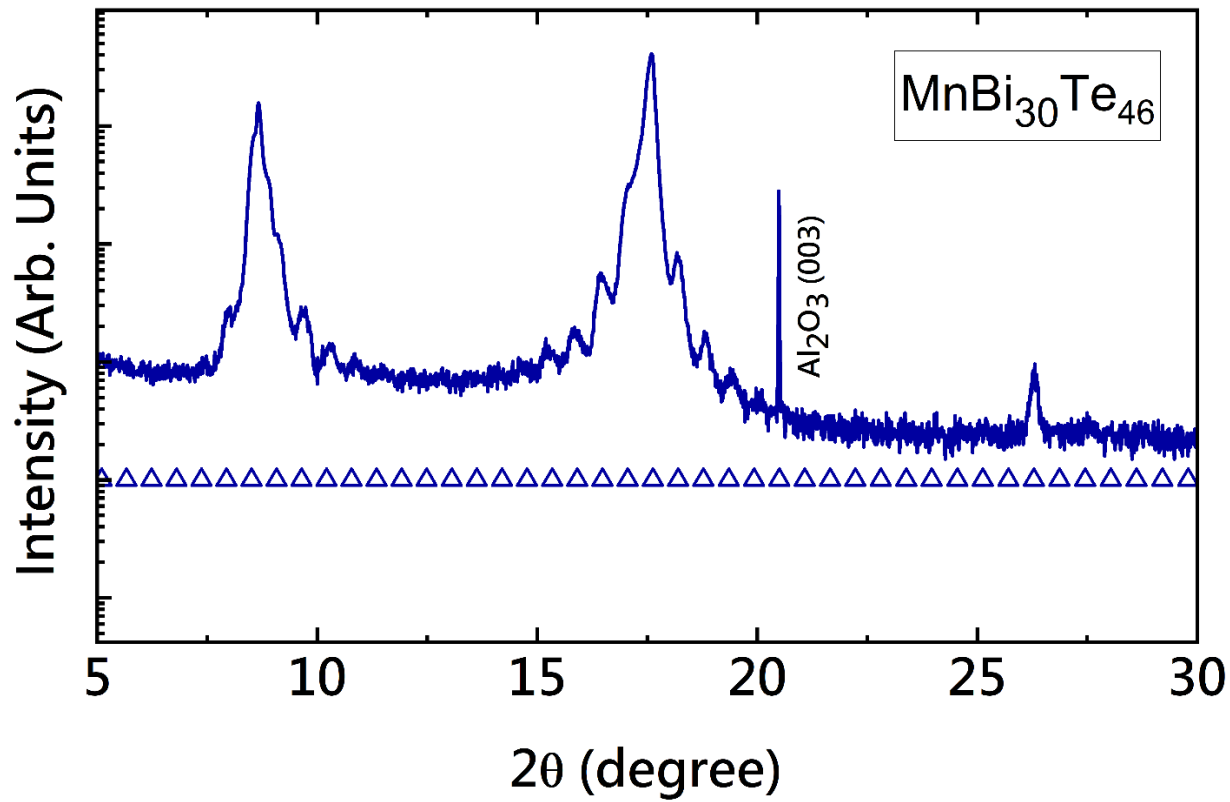

**Figure S1** XRD pattern of [MnTe][Bi$_2$Te$_3$]$_n$ for *n* = 15, with the expected peak values represented by open triangles.

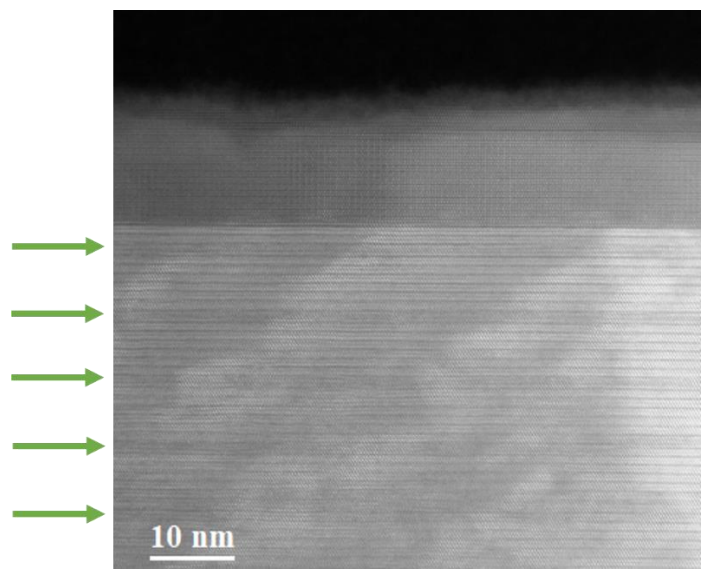

**Figure S2** Cross-sectional HAADF-STEM image of MnBi$_{14}$Te$_{22}$ ($n = 7$), part of which was shown in Figure 2b. The dark septuple layers have been highlighted with arrows.

(a) 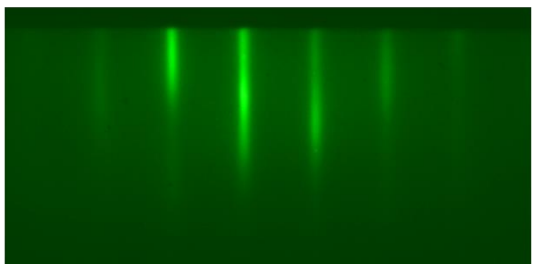

(b) 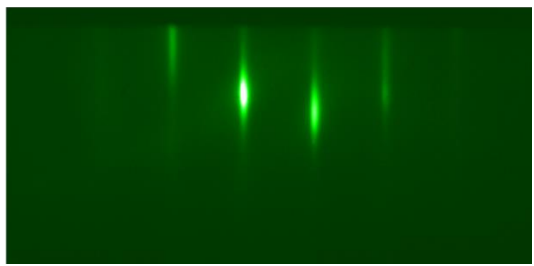

**Figure S3 RHEED images of [MnTe][Bi$_2$Te$_3$]$_n$** (a) $n = 1$ and (b) $n = 5$

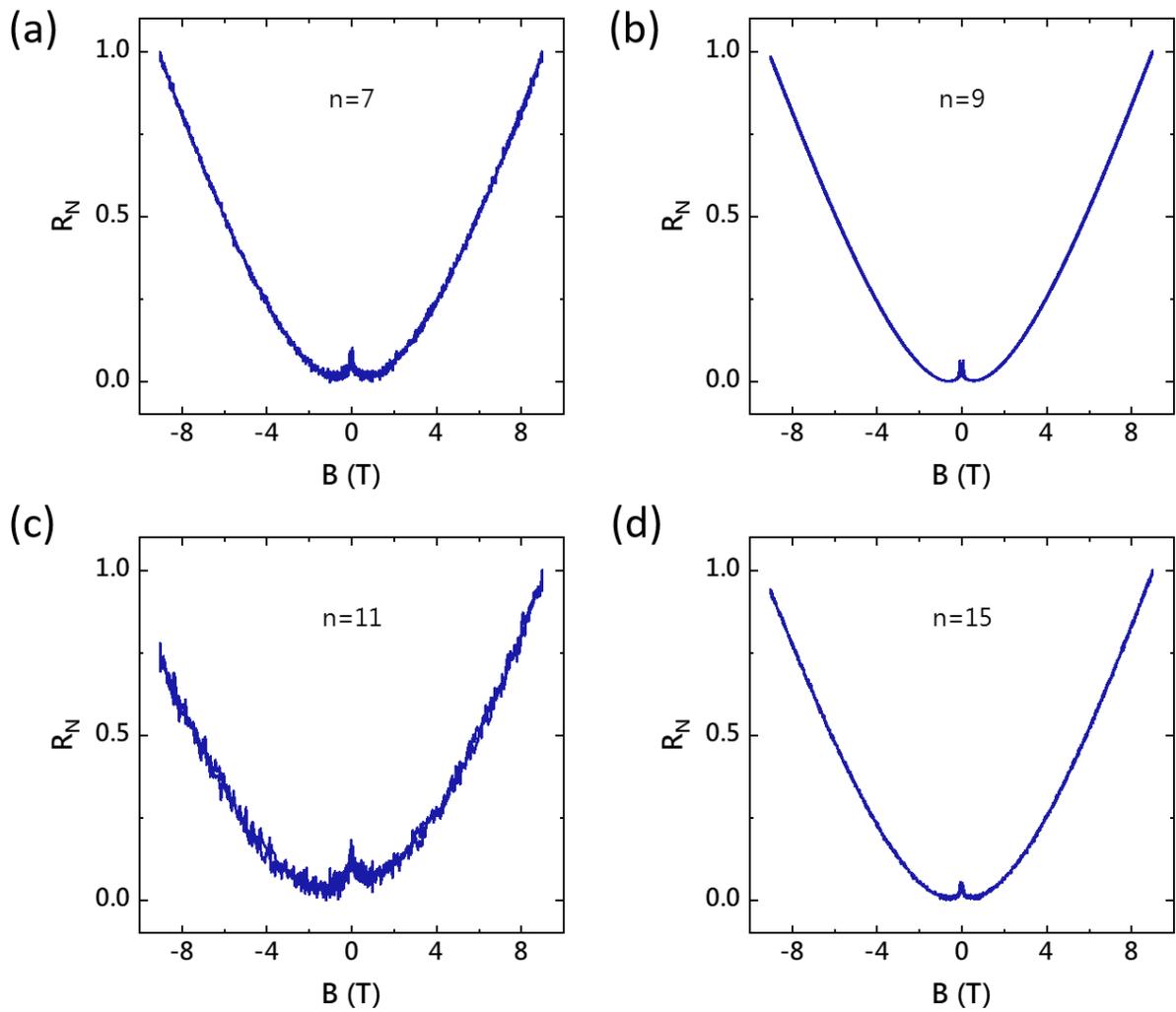

**Figure S4 Magnetoresistance plots of *n* = 7, 9, 11 and 15. (a,b,c,d)** Normalized sheet resistance $R_N = (R-R_{min})/(R_{max}-R_{min})$ for *n* = 7, 9, 11 and 15, respectively.